\documentstyle[12pt]{article}
\input epsf
\def \b{{\cal B}}

\def \cn{Collaboration}

\newcommand{\thspace}{\kern.08333em}

\def \beq{\begin{equation}}
\def \eeq{\end{equation}}
\def \beqn{\begin{eqnarray}}
\def \eeqn{\end{eqnarray}}

\def \v#1#2{V_{#1#2}}

\begin{document}
\rightline{CALT-68-2142}
\rightline{EFI-97-48}
\rightline{hep-ph/9711246}
\rightline{November 1997}
\bigskip
\bigskip
\centerline{{\bf WEAK PHASE $\gamma$ FROM RATIO OF $B \to K \pi$ RATES}
\footnote{Submitted to Phys.~Rev.~D.}}
\bigskip
\centerline{\it Michael Gronau\footnote{Permanent Address: Physics Department,
Technion -- Israel Institute of Technology, 32000 Haifa, Israel.}} 
\centerline{\it California Institute of Technology}
\centerline{\it Pasadena, CA 91125}
\medskip
\centerline{and}
\medskip
\centerline{\it Jonathan L. Rosner}
\centerline{\it Enrico Fermi Institute and Department of Physics}
\centerline{\it University of Chicago, Chicago, IL 60637}
\bigskip
\centerline{\bf ABSTRACT}
\medskip
\begin{quote}
The ratio of partial decay rates for charged and neutral $B$ mesons to $K \pi$
final states provides information on the weak phase $\gamma \equiv {\rm Arg}
(V_{ub}^*)$ when augmented with information on the CP-violating asymmetry in
the $K^\pm \pi^\mp$ mode. The requirements for a useful determination of
$\gamma$ are examined in the light of present information about the decays $B^0
\to K^+ \pi^-$, $B^+ \to K^0 \pi^+$, and the corresponding charge-conjugate
modes. The effects of electroweak penguins and rescattering corrections are
noted, and proposals are made for estimating and measuring their importance. 
\end{quote}
\medskip

\leftline{\qquad PACS codes:  12.15.Hh, 12.15.Ji, 13.25.Hw, 14.40.Nd}
\vfill
\newpage

\centerline{\bf I.  INTRODUCTION}
\bigskip

The leading candidate to describe the violation of CP symmetry in decays of
neutral kaons \cite{CCFT} is the existence of phases in the weak
charge-changing couplings of quarks to $W$ bosons.  These couplings are
parametrized by a unitary $3 \times 3$ matrix, the Cabibbo-Kobayashi-Maskawa or
CKM matrix \cite{CKM} whose elements $V_{ij}$ connect the quarks $i=u,~c,~t$ of
charge 2/3 with those $j=d,~s,~b$ of charge $-1/3$. 

The phase $\gamma = {\rm Arg}(V_{ub}^*)$ (in a standard convention \cite{PDG})
is poorly known in this description.  The unlikely possibility that $\gamma =
0$ would require CP violation in the neutral kaon system to originate elsewhere
than via CKM phases, e.g., via a superweak interaction \cite{WSW}.  Thus, it is
important to seek independent information on $\gamma$, which is provided by the
study of $B$ meson decays. 

Some time ago we proposed a method \cite{DGR} of measuring the weak phases
$\gamma$ and $\alpha$ from the decays $B^0 \to K^+ \pi^-,~B^+ \to K^0
\pi^+,~B^0 \to \pi^+\pi^-$ and from the charge-conjugated processes. In the
present article we explore in detail the part of the method which determines
$\gamma$ utilizing primarily via the ratio of decays of neutral and charged $B$
mesons to $K \pi$ final states \cite{RFa,FM,RFb}.  By combining information on
the charge-averaged ratio 
\beq
R \equiv \frac{\Gamma(B^0 \to K^+ \pi^-) + \Gamma(\bar B^0 \to K^- \pi^+)}
{\Gamma(B^+ \to K^0 \pi^+) + \Gamma(B^- \to \bar K^0 \pi^-)} 
\eeq
with the CP-violating rate asymmetry
\beq
A_0 \equiv \frac{\Gamma(B^0 \to K^+ \pi^-) - \Gamma(\bar B^0 \to 
K^- \pi^+)} {\Gamma(B^+ \to K^0 \pi^+) + \Gamma(B^- \to \bar K^0 \pi^-)}~~~, 
\eeq
we find an expression for $\gamma$ which depends only on these quantities and
on the ratio of tree to penguin amplitudes for which we provide an estimate
based on $B \to \pi \pi$ and $B \to \pi \ell \nu_\ell$ decays.  This method
has become of particular interest now that the CLEO Collaboration has
observed both the $B^0 \to K^+ \pi^-$ and the $B^+ \to K^0 \pi^+$ processes
(and their charge conjugates) \cite{CLEOKpi}.  A similar idea can be applied
to $B_s \to K^+ K^-$ and $B_s \to K^0 \bar K^0$ decays \cite{RFb,DF}.

We define amplitudes and discuss their phases and magnitudes in Section II. The
extraction of $\gamma$ from $B \to K \pi$ decays occupies Section III. Several
potential sources of systematic errors, involving electroweak penguins and
rescattering effects, are studied in Section IV. A generalization of the method
to $B \to K^* \pi$ and $B \to K \rho$ decays is discussed in Section V, and
Section VI concludes. 
\bigskip

\centerline{\bf II.  AMPLITUDES AND THEIR MAGNITUDES}
\bigskip

\leftline{\bf A.  Definitions}
\bigskip

We adopt a flavor-SU(3) decomposition of amplitudes which has been used in
several previous descriptions of $B$ decays to pairs of light pseudoscalar
mesons \cite{DGR,ZepSWChau,GHLR,eta,etap}.  For present purposes the important
amplitudes are strangeness-preserving (unprimed) and strangeness-changing
(primed) amplitudes corresponding to color-favored tree $(T,T')$, penguin
$(P,P')$, and color-suppressed tree $(C,C')$ processes.  The contributions of
electroweak penguins \cite{EWP} may be included by replacing $T \to t \equiv T
+ P_{EW}^C$, $P \to p \equiv P - (1/3)P_{EW}^C$, and $C \to c \equiv C +
P_{EW}$, where the superscript on the electroweak penguin amplitude $P_{EW}$
denotes color-suppression. We stress that, although this general description of
many processes in terms of just a few SU(3) amplitudes assumes flavor SU(3), in
certain cases, such as the one discussed in the subsequent subsection, only
isospin symmetry is required. 

The phases of amplitudes for $\Delta S = 0$ transitions are
Arg($V_{ub}^*V_{ud}) = \gamma$ for tree amplitudes and Arg($V_{tb}^* V_{td}) =
- \beta$ for top-dominated penguin amplitudes.  For $|\Delta S| = 1$ the
corresponding phases are Arg($V_{ub}^*V_{us}) = \gamma$ (tree) and
Arg($V_{tb}^* V_{ts}) = \pi$ (top-dominated penguin). Nothing changes in the
$|\Delta S| = 1$ penguin transitions if these receive important contributions
from $c \bar c$ intermediate states \cite{BF,Ciu}. While the phase of the
$\Delta S = 0$ penguin amplitude may be affected under such circumstances
\cite{BF}, we shall not be concerned with this phase. 

In the rest of this Section we ignore amplitudes which involve the
participation of the spectator quark. Without rescattering, these amplitudes
are expected to be suppressed by a factor of $f_B/m_B$, where $f_B \sim 200$
MeV is the $B$ meson decay constant. The neglect of these contributions was
noted to be equivalent to an assumption that some rescattering effects are
unimportant, thus leading to the vanishing of certain final state interaction
phase-differences \cite{GHLR,LINC,GLR,GEWY}. Such amplitudes can also be
generated by rescattering from intermediate states obtained by
$T^{(')},~P^{(')},~C^{(')}$ amplitudes. Using a Regge analysis to demonstrate
rescattering effects \cite{DGPS,BH}, it was shown \cite{BGR} that such
amplitudes may be suppressed only by a factor of about 0.2  \cite{RESCAT}
rather than by $f_B/m_B\sim 0.04$, in which case explicit tests for such
rescattering can be performed. The effect of these rescattering amplitudes,
assumed to be as large as estimated in Ref.~\cite{BGR}, will be studied in
Sec. IV B. Similarly, we begin by ignoring electroweak penguin contributions,
deferring their treatment to Secs.~IV A. 
\bigskip

\leftline{\bf B.  Decomposition for $B \to K \pi$ decays}
\bigskip

The amplitude for $B^+ \to K^0 \pi^+$ is given by a QCD-penguin contribution: 
\beq \label{eqn:B+}
A(B^+ \to K^0 \pi^+) = - |P'|~~~,
\eeq
where we have adopted a convention in which all strong phases are expressed
relative to that in the $|\Delta S| = 1$ penguin amplitude. As a consequence,
one expects no CP-violating difference between the partial widths $\Gamma_{0+}
\equiv \Gamma(B^+ \to K^0 \pi^+)$ and $\Gamma_{0-} \equiv \Gamma(B^- \to \bar
K^0 \pi^-)$.  (We use a notation in which the subscripts denote the charges of
the final kaon and pion.)  For brevity we shall thus define $\Gamma_C \equiv
\Gamma_{0+} = \Gamma_{0-}$ to be the partial width for a charged $B$ to decay
to a neutral kaon and a charged pion. 

The amplitude for $B^0 \to K^+ \pi^-$ is expected to be dominated by the
penguin contribution $P'$. One only uses isospin symmetry to relate the penguin
amplitudes in neutral and charged $B$ decays to $K \pi$ states. The tree
contribution can be roughly estimated \cite{GHLR}, $|T'/P'| \sim 0.2$. We shall
refine this estimate presently.  Thus 
\beq \label{eqn:B0}
A(B^0 \to K^+ \pi^-) = |P'| - |T'|e^{i \delta} e^{i \gamma}~~,~~~
A(\bar B^0 \to K^- \pi^+) = |P'| - |T'|e^{i \delta} e^{-i \gamma}~~~,
\eeq
where $\delta$ is the strong phase difference between the tree and penguin
amplitudes, and we have used the weak phases given in Table 1. The
corresponding rates may be defined as $\Gamma_{+-} \equiv \Gamma(B^0 \to K^+
\pi^-)$ and $\Gamma_{-+} \equiv \Gamma(\bar B^0 \to K^- \pi^+)$. A CP-violating
rate asymmetry $\Gamma_{+-} \ne \Gamma_{-+}$ may arise whenever both $\sin
\delta$ and $\sin \gamma$ are nonvanishing.  At the same time, even if $\sin
\delta = 0$ so that $\Gamma_{+-} = \Gamma_{-+}$, these two partial rates may
differ from $\Gamma_C$ as a result of the extra $T'$ contribution they contain
\cite{RFa,FM,RFb}.  This difference can shed light on the weak phase $\gamma$.
Our purpose in the present paper is to estimate the experimental demands on
such a determination. 
\bigskip

\leftline{\bf C.  Magnitudes of amplitudes}
\bigskip

The CLEO Collaboration \cite{CLEOKpi} has observed both $B^0 \to K^+ \pi^-$ and
$B^+ \to K^0 \pi^+$ (here we do not distinguish between a process and its
charge conjugate).  The observed branching ratios are
\beq \label{eqn:neut}
\b(B^0 \to K^+ \pi^-) = (15^{+5}_{-4} \pm 1 \pm 1) \times 10^{-6}~~~,
\eeq
\beq \label{eqn:chgd}
\b(B^+ \to K^0 \pi^+) = (23^{+11}_{-10} \pm 3 \pm 2) \times 10^{-6}~~~.
\eeq
We shall express squares of amplitudes in units of $B$ branching ratios times
$10^6$. In Ref.~\cite{etap} we averaged the rates (\ref{eqn:neut}) and
(\ref{eqn:chgd}) to obtain $|P'|^2 = 16.3 \pm 4.3$.  However, here we shall
leave open the possibility that a significant $T'$ contribution is affecting
the $B^0 \to K^+ \pi^-$ rate, and take $|P'|^2 = 23 \pm 10.5$ from the $B^+ \to
K^0 \pi^+$ rate. 

We estimate $|T'|$ by relating it through flavor SU(3) to the corresponding
strangeness-preserving amplitude $|T|$ governing such decays as $B^0 \to \pi^+
\pi^-$ and $B^+ \to \pi^+ \pi^0$.  Using the phase conventions of
Ref.~\cite{GHLR}, we find 
\beq
A(B^0 \to \pi^+ \pi^-) = -(T + P)~~;~~~
A(B^+ \to \pi^+ \pi^0) = -(T + C)/\sqrt{2}~~~.
\eeq
Although neither process has been observed with a statistically significant
signal, Ref.~\cite{CLEOKpi} quotes a $2.8 \sigma$ signal of
\beq \label{eqn:ppz}
\b(B^+ \to \pi^+ \pi^0) = (9^{+6}_{-5}) \times 10^{-6}
\eeq
and a $2.2 \sigma$ signal of
\beq \label{eqn:pp}
\b(B^0 \to \pi^+ \pi^-) = (7 \pm 4) \times 10^{-6}~~~.
\eeq
Taking (\ref{eqn:ppz}) as an estimate of $|T|^2/2 = 9 \pm 5.5$ (neglecting the
color-suppressed amplitude $C$ in $B^+ \to \pi^+ \pi^0$), and (\ref{eqn:pp}) as
an estimate of $|T|^2 = 7 \pm 4$ (neglecting the penguin amplitude $P$ in $B^0
\to \pi^+ \pi^-$), we find \cite{etap} that $|T|^2 = 8.3 \pm 3.8$.  The
observed rate for the semileptonic decay $B^0 \to \pi^- \ell^+ \nu_\ell$
\cite{semi}, 
\beq \label{eqn:sl}
\b(B^0 \to \pi^- \ell^+ \nu_\ell) = (1.8 \pm 0.4 \pm 0.3 \pm 0.2) \times
10^{-4}~~~,
\eeq
is compatible with this estimate if one calculates the $B^0 \to \pi^+ \pi^-$
decay via factorization.  An early estimate \cite{Vol} based on a form factor
dominated by the $B^*$ pole, 
\beq
\frac{\Gamma(B^0 \to \pi^- \ell^+ \nu_\ell)}{\Gamma(B^0 \to \pi^+ \pi^-)} =
\frac{M_B^2}{12 \pi^2 f_\pi^2} \simeq 13~~~,
\eeq
would imply $\b(B^0 \to \pi^+ \pi^-) = (1.4 \pm 0.4) \times 10^{-5}$ and hence
$|T|^2 = 14 \pm 4$ on the basis of the observed semileptonic rate
(\ref{eqn:sl}). More recent estimates \cite{LGsl} yield a similar range of
values. An improvement of the data will allow one to focus on the $q^2$ value
appropriate to pion (or kaon) production and thus to reduce the dependence on
models drastically. The direct CLEO upper limit \cite{CLEOKpi} $\b(B^0 \to
\pi^+ \pi^-) < 1.5 \times 10^{-5}$ (90\% c.l.) gives a poorer upper limit on
$|T|^2$ than our estimate. 

One then uses factorization which introduces SU(3) breaking through a factor
$f_K/f_{\pi}$ \cite{GHLR} to predict 
\beq \label{eqn:tpt}
|T'/T| = (f_K/f_\pi) |V_{us}/V_{ud}| = 0.27
\eeq
with an error estimated to be about 20\% (the typical breaking of flavor SU(3)
symmetry) \cite{Gib}. Combining the estimates for amplitudes and their ratio,
we then find 
\beq
r \equiv |T'/P'| = 0.16 \pm 0.06~~~.
\eeq

The estimate of $|T'|$ is likely to improve in the future once the
spectrum for the semileptonic decay $B^0 \to \pi^- \ell^+ \nu_\ell$ is measured
at $q^2 = m_K^2$.  One uses factorization directly \cite{Bj,BS} to predict
\beq \label{eqn:dsl}
\Gamma(B^0 \to K^+ \pi^-)|_{\rm tree} = 6 \pi^2 f_K^2 |V_{us}|^2 a_1^2
\frac{d \Gamma(B^0 \to \pi^- \ell^+ \nu_\ell)}{dq^2} |_{q^2 = m_K^2}~~~,
\eeq
with $a_1 = 1.08 \pm 0.04$ \cite{Neubert}.  This value was obtained from a fit
to $b \to c \bar u d$ subprocesses in $B \to D^{(*)} + \pi(\rho)$ decays.  The
value appropriate to the subprocesses $b \to u \bar u d$ and $b \to u \bar u s$
which contribute to the tree amplitudes in $B \to \pi \pi$ and $B \to K \pi$
may be slightly different.  It may be difficult to determine $a_1$ to an
accuracy of better than 10\% in these processes as a result of penguin ($P$)
amplitudes accompanying the factorizable color-allowed ($T \sim a_1$) and
non-factorizable color-suppressed $(C \sim a_2)$ $\Delta S = 0$ amplitudes. (In
contrast, $B \to D \pi$ has no penguin contributions.) 

Since the present branching ratio (\ref{eqn:sl}) is known to about 30\%, a
factor of 100 increase in the data sample (envisioned in future high-intensity
studies) would permit this branching ratio to be known to about 3\%.  More
crucial is the error on the differential rate on the right-hand side of
Eq.~(\ref{eqn:dsl}).  As we shall see, a 10\% determination of $|T'|$ (hence a
20\% accuracy in the differential rate at $q^2 = m_K^2$) is the accuracy that
will be required in order for the present method to be reasonably useful. 
\bigskip

\centerline{\bf III.  EXTRACTION OF $\gamma$ FROM $B \to K \pi$ RATES}
\bigskip

\leftline{\bf A.  The Fleischer-Mannel bound}
\bigskip

Recalling the definitions of Sec.~II, we may form the ratio
\beq
R \equiv \frac{\Gamma(B^0 \to K^+ \pi^-) + \Gamma(\bar B^0 \to K^- \pi^+)}
{\Gamma(B^+ \to K^0 \pi^+) + \Gamma(B^- \to \bar K^0 \pi^-)}
= \frac{\Gamma_{+-} + \Gamma_{-+}}{2 \Gamma_C}~~~
\eeq
which has the simple form \cite{RFa,FM,RFb}
\beq \label{eqn:bigR}
R = 1 - 2 r \cos \gamma \cos \delta + r^2~~~.
\eeq
Fleischer and Mannel \cite{FM} have pointed out that if $R < 1$ a useful bound
on $\gamma$ can be obtained regardless of the value of $r$ or $\delta$: 
\beq \label{eqn:FM}
\sin^2 \gamma \leq R~~~.
\eeq
The present value of $R$ is $0.65 \pm 0.40$, so a reduction of errors by a
factor of 3 with no change in central value would begin to provide a useful
limit excluding some region around $\gamma = \pi/2$. 
\bigskip

\leftline{\bf B.  Use of information on $r$}
\bigskip

In the presence of information on $r$ (see also Ref.~\cite{RFb}) one can
provide a more precise estimate of $\gamma$ by measuring the difference in $B^0
\to K^+ \pi^-$ and $\bar B^0 \to K^- \pi^+$ decay rates.  One forms the
pseudo-asymmetry 
\beq 
A_0 \equiv \frac{\Gamma(B^0 \to K^+ \pi^-) - \Gamma(\bar B^0 \to K^- 
\pi^+)}
{\Gamma(B^+ \to K^0 \pi^+) + \Gamma(B^- \to \bar K^0 \pi^-)}
= \frac{\Gamma_{+-} - \Gamma_{-+}}{2 \Gamma_C}~~~.
\eeq
Note that the denominator is taken to be $2 \Gamma_C$ in order to divide
by $|P'|^2$ without any complication from the $T'$ amplitude.  Since 
\beq \label{eqn:Ap}
A_0 =  2 r \sin \delta \sin \gamma~~~,
\eeq
one can combine (\ref{eqn:bigR}) and
(\ref{eqn:Ap}) to eliminate $\delta$.  The result is
\beq \label{eqn:Req}
R = 1 + r^2 \pm \sqrt{4 r^2 \cos^2 \gamma - A_0^2 \cot^2 \gamma}~~~.
\eeq
This quantity is plotted in Fig.~1 for $r = 0.16$ (the central value of our
estimate) and various values of $A_0$.  (Note that the results are
insensitive to the sign of $A_0$.)  We vary $r$ between its limits and
plot $R$ for $r = 0.10$ in Fig.~2 and for $r = 0.22$ in Fig.~3. 

\begin{figure}
\centerline{\epsfysize = 4 in \epsffile {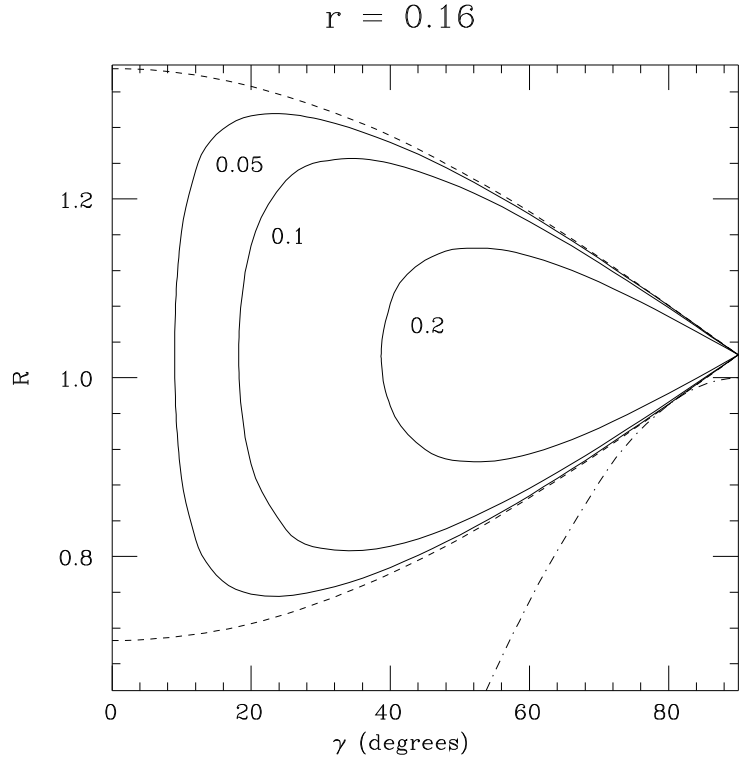}}
\caption{Value of $R$ (ratio of neutral to charged $B \to K \pi$ partial
widths) as a function of $\gamma = {\rm Arg}(V^*_{ub})$ for $r=0.16$.  Solid
lines are labeled by values of pseudo-asymmetry parameter $|A_0|$.  Dotted
boundary lines correspond to $A_0=0$.  The case $\gamma = 0$ for arbitrary $R$
between the bounds of the dashed lines also corresponds to $A_0=0$.  Also shown
(dot-dashed lines) is the Fleischer-Mannel bound $\sin^2 \gamma \leq R$.} 
\end{figure}

\begin{figure}
\centerline{\epsfysize = 3 in \epsffile {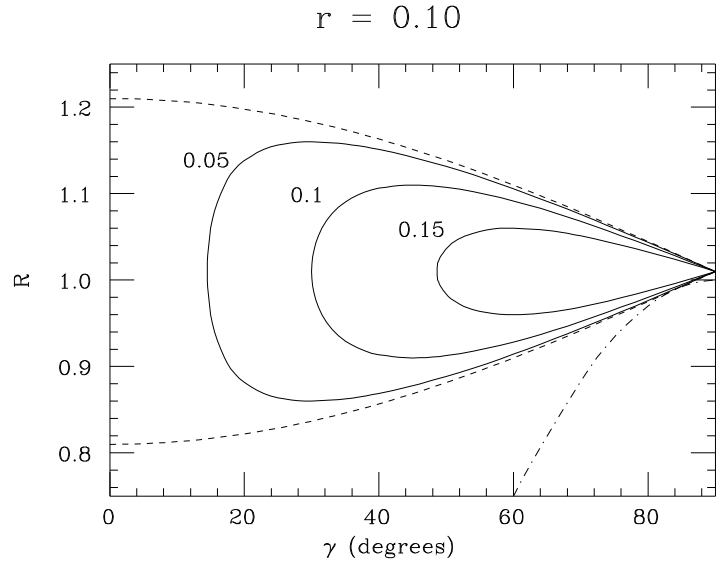}}
\caption{Same as Fig.~1 but for $r=0.10$.}
\end{figure}

Let us assume for the moment that $r$ is known.  Figs.~1--3 have several
interesting features. 

\begin{enumerate}

\item The maximum value of $|A_0| = 2r$ occurs only for $\gamma = \delta =
90^{\circ}$ (as one sees from (\ref{eqn:Ap}).

\item The result is symmetric with respect to $\gamma \to \pi - \gamma$, since
$A_0$ is only sensitive to $\sin \gamma$ and the $\cos \delta$ term in $R$
involves a sign arbitrariness. 

\item The sensitivity to $\gamma$ in the range $\gamma > 45^{\circ}$ is
greatest for $A_0=0$. 

\item As long as $A_0 \ne 0$ there will be two solutions for $\gamma$ in the
range $0 \le \gamma \le \pi/2$ (and two in the range $\pi/2 \le \gamma \le
\pi$) for any given $R$. Observation of a non-zero $A_0$ would rule out the
possibility of $\gamma = 0$ mentioned in the introduction and hence would
disprove a superweak model of CP violation.

\end{enumerate}

The sign of $\cos\delta$, which would resolve the ambiguity between $\gamma$
and $\pi - \gamma$, can be studied theoretically. One model-dependent
calculation \cite{GH} finds $\delta < \pi/2$, implying that $R$ is smaller
(larger) than $1 + r^2$ for $\gamma$ smaller (larger) than $\pi/2$. This model
calculation ignores possible phases due to soft final state interactions
\cite{DGPS,WOLF}. 

Eq.~(\ref{eqn:Req}) can be inverted to obtain a quadratic equation for $\sin^2
\gamma$ in terms of $r$, $A_0$, and $R$.  We then find 
$$
4r \sin \gamma = \pm \{ [(1+r)^2-(R+A_0)][(R-A_0)-(1-r)^2] \}^{1/2}
$$
\beq \label{eqn:gamma}
\pm \{ [(1+r)^2-(R-A_0)][(R+A_0)-(1-r)^2] \}^{1/2}~~~. 
\eeq
This relation also follows directly from the geometry of a triangle formed by
the amplitudes $A(B^+ \to K^0 \pi^+),~T',~A(B^0 \to K^+ \pi^-)$ and the
charge-conjugated triangle. 
\begin{figure}
\centerline{\epsfysize = 5.25 in \epsffile {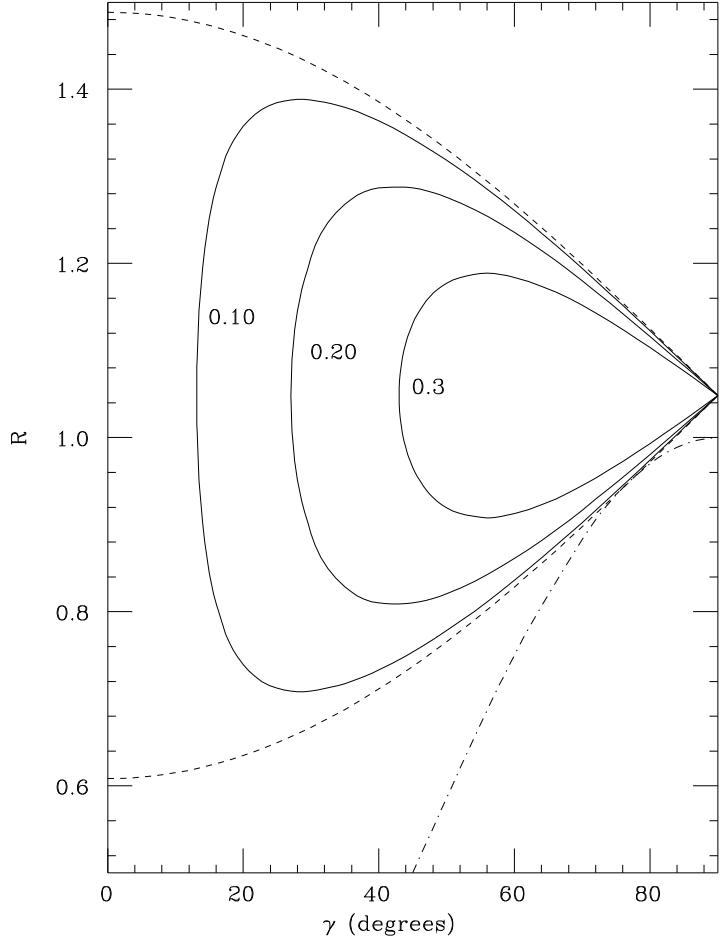}}
\caption{Same as Fig.~1 but for $r=0.22$.}
\end{figure}
The triangle construction is similar to one employed in Ref.~\cite{GW} for
obtaining $\gamma$ from $B \to D K$ decays.  One measures the sides of two
triangles (for three processes and their charge conjugates), sharing a common
base ($P'$ in the present case), where another pair of sides in both triangles
is equal in length ($T'$ in the present case), and forms an angle $2 \gamma$. 
Just as in the $B \to DK$ case, one of the sides of each triangle $(T')$ is
much smaller than the others. This leads to large experimental uncertainties in
determining $\gamma$, following from relatively small experimental errors in
the side measurements (square root of rates).  Furthermore, the magnitude of
the small $T'$ suffers the largest theoretical uncertainty. So, if one draws
these triangles in roughly correct proportions, one can see why the uncertainty
in $\gamma$ is likely to be large. 
\bigskip

\leftline{\bf C.  Required precision}
\bigskip

The precision in $R$, $r$, and $A_0$ required to measure $\gamma$ to a given
level depends on their values.  We consider two extreme cases of final state
phases which bracket others. In the first, with $\sin \delta \simeq 0$,
corresponding to parameters near the dashed boundary curves in Figs.~1--3, the
error in $\gamma$ is dominated by the errors in $R$ and $r$. In the second,
corresponding to parameters midway between the dashed boundary curves in
Figs.~1--3, with $\cos \delta \simeq 0$ and $R \simeq 1 + r^2$, the error in
$\gamma$ is due to the errors in $r$ and $A_0$. 

(1) When $\sin \delta = 0$, we have the simple result
\beq
\cos \gamma = \pm \frac{R - 1 - r^2}{2r}~~~,
\eeq
so that
\beq
\left. \frac{\partial \gamma }{\partial R} \right|_r = \mp \frac{1}{2 r \sin
 \gamma}~~,~~~
\left. \frac{\partial \gamma}{\partial r} \right|_R = \frac{\cos \gamma \pm r}
{r \sin \gamma}~~~.
\eeq
Let us take as an example the case $r = 0.16$, $A_0 = 0$, and $\gamma =
90^\circ$.  Then $\Delta \gamma = 5^\circ$ corresponds to $|\Delta R| = 0.028$
for fixed $r$ and $|\Delta r| = 0.087$ for fixed $R$.  The requirement on $R$
is more stringent.  In order to satisfy it, we need about 2500 events in both
charged and neutral modes of $B \to K \pi$, or roughly 200 times the present
sample. This is thought to be within the capabilities of an upgraded version of
CESR \cite{AW}, as well as dedicated hadronic $B$ production experiments at the
Fermilab Tevatron and the CERN Large Hadron Collider \cite{Stone}. 

A second example with $\sin \delta= 0$ exhibits greater sensitivity to $r$. 
When $\gamma = 45^\circ$ or $135^\circ$ (which is about as far from $\gamma =
90^\circ$ as allowed by present fits \cite{Cargese}) we have 
\beq
\left. \frac{\partial \gamma}{\partial R} \right|_r = \mp \frac{1}{\sqrt{2} r}
~~,~~~ \left. \frac{\partial \gamma}{\partial r} \right|_R = \pm \sqrt{2}
+ \frac{1}{r}~~~.
\eeq
Choosing $r=0.16$ and the postive sign in the second of the above two equations
to exhibit the more stringent requirement, we find that an error of $|\Delta R|
= 0.028$, which as noted in the previous example is thought to be within reach
in future experiments, corresponds to $\Delta \gamma = 7^\circ$.  An error in
$\gamma$ of this same magnitude (for $r=0.16$) is associated with $\Delta r =
0.016$, i.e., a 10\% error in $r$ or a 20\% error on the quantity $d \Gamma(B^0
\to \pi^- \ell^+ \nu_\ell)/ dq^2$ at $q^2 = m_K^2$, as noted at the end of
Sec.~II C. Thus, one can envision determining $\gamma$ to an overall error of
$[(7^\circ)^2 + (7^\circ)^2]^{1/2} \simeq 10^\circ$ if $\gamma = 45^\circ$ or
$135^\circ$, and if the required precision on $r$ can be achieved. 

(2) When $\cos \delta = 0$, the magnitude of the asymmetry $|A_0|$ is just
$2 r \sin \gamma$, so
\beq
\left. \frac{\partial \gamma}{\partial |A_0|} \right|_r =  \frac{1}{2 r
\cos \gamma}~~,~~~
\left. \frac{\partial \gamma}{\partial r} \right|_{|A_0|} = - \frac{\tan
\gamma}{r}~~~.
\eeq
Taking $\gamma = 45^{\circ}$ and $r = 0.16$, we find $\Delta \gamma =
7^{\circ}$ corresponds to $\Delta |A_0| = 0.028$ for fixed $r$ and $\Delta r
= 0.02$ for fixed $|A_0|$.  Measurement of $|A_0|$ to $\pm 0.028$ requires
1250 events each of $B^0 \to K^+ \pi^-$ and $\bar B^0 \to K^- \pi^+$.  Thus,
a measurement of $\gamma$ with an overall error of less than $10^\circ$
appears feasible in this case as well.

The reduction of $\Delta r/r$ by about a factor of four to $\simeq 10\%$
appears to be at the limits of understanding of form factors which would permit
determination of $T$ in $B \to \pi^+ \pi^-$ from factorization and $B \to \pi
\ell \nu_\ell$.  (The $B \to \pi \ell \nu$ decay is claimed to be capable of
yielding an accuracy of 10\% in determining $|V_{ub}|$ \cite{Ball}.)  As noted
at the end of Sec.~II C, if factorization is found to be reliable in comparing
$B^0 \to \pi^+ \pi^-$ and $B \to \pi \ell \nu$ decays, one may be able to pass
directly from $B \to \pi \ell \nu$ decays to an estimate of the tree ($T'$)
contribution to $B^0 \to K^+ \pi^-$, since all that is required is for the weak
current to produce a kaon, a process which we can estimate reliably. 
\bigskip

\centerline{\bf IV. SYSTEMATIC ERRORS}
\bigskip

Aside from the statistical errors analyzed in Sec.~III, we have noted that a
systematic theoretical error in determining $r$ of $\simeq 10\%$ seems to be
unavoidable.  An error of this magnitude is encountered whether we determine
$r$ from $B^0 \to \pi^+ \pi^-$ and $B^+ \to \pi^+ \pi^0$ decays, thereby
omitting nonfactorizable $P$ and $C$ terms, respectively, or from $B \to \pi
\ell \nu_\ell$ decays. In addition, two smaller contributions to the amplitudes
should be noted: (A) the effects of the color-suppressed electroweak penguin
terms $P'^C_{EW}$ in $B^0 \to K^+ \pi^-$ and $B^+ \to K^0 \pi^+$, and (B) the
effect of the annihilation amplitude (called $A'$ in Refs.~\cite{GHLR}) (or
rescattering effects) on $B^+ \to K^0 \pi^+$. 
  
Fleischer and Mannel find a very small electroweak penguin term \cite{FM},
$|P'^C_{EW}/P'| < 0.01$. The small value is obtained as a result of a delicate
cancellation among larger contributions from electroweak penguin operators
which have a different color structure. The calculation is based on
factorization of hadronic matrix elements of QCD and electroweak penguin
operators.  With all uncertainties involved, a conservative estimate should
allow values of $|P'^C_{EW}/P'|$ at a level of 5\% \cite{FMb}, given roughly by
a product of the ratio of corresponding Wilson coefficients and a color factor
\cite{GHLR}. The effect of electroweak penguin contributions on our analysis
will be studied in subsection A. We will also suggest ways of measuring
electroweak penguin contributions in related processes. 

A small $A'$ term should also be allowed in $B^+ \to K^0 \pi^+$. A naive
estimate of $A'$ neglecting rescattering, $|A'/T'| \sim f_B/m_B$, yields
$|A'/P'|\sim 0.01$. (A similar estimate applies to the $u$-quark contribution
to $P'$.) Rescattering effects are hard to calculate. Regge-model estimates
\cite{BH,BGR}, in which rescattering from intermediate states such as
$K^+\pi^0$ is described by a $\rho$-trajectory exchange, suggest $|A'/T'| \sim
0.2$. The consequences of such rescattering effects will be studied in
subsection B. A few methods for direct measurements of rescattering effects in
SU(3)-related processes will also be described. 
\bigskip

\leftline{\bf A. Modification due to electroweak penguins}
\bigskip

In the presence of electroweak penguin contributions, which carry the same weak
phase as $P'$, Eqs.~(\ref{eqn:B+}) and (\ref{eqn:B0}) take the form
\cite{DGR} 
\beq
A(B^+ \to K^0 \pi^+) = A(B^- \to \bar K^0 \pi^-) = - |p'|~~,~~~p' \equiv P'
-(1/3)P'^C_{EW}~~~,
\eeq
\beq
A(B^0 \to K^+ \pi^-) = |p' + P'^C_{EW}| - |T'|e^{i \delta} e^{i \gamma}~~,~~~
A(\bar B^0 \to K^- \pi^+) = |p' + P'^C_{EW}| - |T'|e^{i \delta} e^{-i 
\gamma}~~~.
\eeq
A common unmeasurable strong phase in the $B^0$ and $\bar B^0$ decay amplitudes
has been omitted; $\delta$ is the corrected strong phase difference.
Consequently, the expressions for $R$ and $A_0$ are modified as follows: 
\beq
R/a^2 = 1 - 2 r' \cos \gamma \cos \delta + r'^2~~,~~~
A_0/a^2 =  2 r' \sin \delta \sin \gamma~~~,
\eeq
where $a = |1 + (P'^C_{EW}/p')|$ and $r' = (1/a)|T'/p'| = |T'/(p' +
P'^C_{EW})|$. The Fleischer-Mannel bound becomes $\sin^2 \gamma \leq R/a^2$.

Using $|P'^C_{EW}/p'| < 0.05$, and assuming an arbitrary relative strong
phase difference between $p'$ and $P^C_{EW}$, one has $a^2 = 1.0 \pm 0.1$. This
factor normalizes both the ratio of rates $R$ and the asymmetry $A_0$. The
electroweak penguin terms introduce an additional 5\% uncertainty in the ratio
of tree-to-penguin amplitudes $r'$. 

An important question is whether $a$ is larger or smaller than one.
Model-dependent perturbative calculations \cite{FM,GH} of QCD and electroweak
penguin amplitudes suggest that the strong phase difference between $p'$ and
$P'^C_{EW}$ is smaller than $\pi/2$, hence $a > 1$. This would imply that both
$R$ and $A_0$ can only be increased by electroweak penguin contributions and
that the Fleischer-Mannel bound is maintained. Since such calculations
disregard possible phases due to soft final state interaction \cite{DGPS,WOLF},
one cannot exclude, however, the possibility that $a < 1$. 

One way to obtain a clue to the sign of $a - 1$ is to compare the rates for
$B^+ \to K^0 \pi^+$ and $B^0 \to K^0 \pi^0$: 
\cite{GHLR}
\beq \label{eqn:ratio}
\frac{2\Gamma(B^0 \to K^0 \pi^0)}{\Gamma(B^+ \to K^0 \pi^+)} = |1 -
\frac{P'_{EW}}{p'}|^2~~~,
\eeq
where a smaller color-suppressed tree ($C'$) term was neglected. A measurement
of this ratio would determine whether the relative strong phase between
$P'_{EW}$ and $p'$ is larger or smaller than $\pi/2$. Although, in principle,
$P'_{EW}$ and $P'^C_{EW}$ can carry different strong phases, it seems likely
that this information would be sufficient to determine whether $a$ is larger or
smaller than one. The deviation of the ratio (\ref{eqn:ratio}) from one would
also provide some information on the magnitude of the color-allowed electroweak
amplitude $P'_{EW}$, which could provide a useful measure for the smaller
$P'^C_{EW}$ term.  Other ways of measuring the importance of electroweak
penguins have been noted in Refs.~\cite{EWP,EWPmeas}.

The inclusion of electroweak penguins in Fig.~1$-$3 is straightforward. The
figures are to be interpreted as plots of $R/a^2$ versus $\gamma$ for different
values of $A_0/a^2$ and for a fixed value of $r'$. The $1/a^2$ factor
involves a 10\% uncertainty. Such an uncertainty in $R/a^2$ is seen to lead to
a rather large theoretical error in determining $\gamma$, typically of a few
tens of degrees. The error decreases with increasing $r'$. 
\bigskip

\leftline{\bf B. Modification due to rescattering}
\bigskip

In the presence of final state rescattering, the general decomposition of the
decay amplitudes of charged and neutral $B$ mesons is given in terms of
amplitudes carrying specific weak phases \cite{GHLR}:
\beq
A(B^+ \to K^0 \pi^+) = - |p'| + |A'|e^{i \Delta} e^{i \gamma}~~~,
\eeq
\beq
A(B^0 \to K^+ \pi^-) = |p' + P'^C_{EW}| - |T'|e^{i \delta} e^{i \gamma}~~~.
\eeq
We will assume that the magnitude of $A'$ (acquiring an unknown strong phase
$\Delta$), dominated by rescattering from intermediate states such as
$K^+\pi^0$ and multibody states, is given by $|A'/T'| \sim 0.2$. Estimates
of the other ratios of amplitudes were given in previous sections, $r' \approx
|T'/p'| \sim 0.2$ and $|P'^C_{EW}/p'| \sim 0.05$. 

The general expressions for $R$ and $A_0$ are
\beq
(\frac{f}{a^2})R = 1 - 2 r' \cos \gamma \cos \delta + r'^2~~,~~~
(\frac{f}{a^2})A_0 =  2 r' \sin \delta \sin \gamma~~~,
\eeq 
where $f \equiv 1 - 2|A'/p'| \cos \Delta \cos \gamma + |A'/p'|^2$. The
interference between $p'$ and $A'$ in $B^+ \to K^0\pi^+$ leads to an asymmetry 
\beq
A_+ \equiv \frac{\Gamma(B^+\to K^0\pi^+) - \Gamma(B^- 
\to\bar{K}^0\pi^-)}
{\Gamma(B^+\to K^0\pi^+) + \Gamma(B^- \to\bar{K}^0\pi^-)}~~~,
\eeq
which is given by $f A_+ = 2|A'/p'| \sin \Delta \sin \gamma$. Since $|A'/p'|
\sim 0.05$, this asymmetry is not expected to exceed the level of 10$\%$. 

The factor $f/a^2$ in $R$ and $A_0$ involves roughly equal hadronic
uncertainties due to rescattering ($f$) and electroweak penguin ($a^2$)
contributions. Although it is unlikely that this factor differs from unity by
more than 20$\%$ ($a^2/f = 1 \pm 0.2$), it introduces sizable uncertainties in
the determination of $\gamma$ as described in Sec. III. The Fleischer-Mannel
bound becomes $\sin^2 \gamma \leq (f/a^2)R$. The usefulness of this method in
determining (or at least constraining) $\gamma$ depends crucially on future
experimental limits on rescattering effects. Let us mention a few such possible
measurements in SU(3)-related processes. 

The most direct measurements of rescattering effects can be made in $B$ decay
processes, which in the framework of a diagramatic SU(3) description
\cite{GHLR} proceed only through annihilation of the $b$ and spectator quarks.
Such decays may also proceed via  rescattering from other less-suppressed
amplitudes. In this case, Regge-model estimates \cite{BH,BGR} suggest that
ratios of amplitudes such as $|A'/T'|$ and $|E/T|$ are enhanced from
$f_B/m_B\sim 0.04$ (without rescattering) to $\sim 0.2$ (with rescattering). A
list of all such processes, in which a $B$ meson decays to two pseudoscalars is
given in Ref.~\cite{BGR}. 

Consider, for instance, $B^0 \to K^+ K^-$ which is given by the amplitude $E$.
In the absence of rescattering one expects $\b(B^0 \to K^+ K^-)/\b(B^0 \to
\pi^+ \pi^-) \approx |E/T|^2\sim (f_B/m_B)^2 \sim 0.002$. [SU(3) breaking and
the penguin contribution to $B^0\to\pi^+\pi^-$ are neglected.] For $\b(B^0
\to\pi^+ \pi^-) = 10^{-5}$, this would imply $\b(B^0 \to K^+ K^-) \sim 2 \times
10^{-8}$, or 200 times smaller than the present CLEO upper limit
\cite{CLEOKpi}. On the other hand, a description of rescattering into $K^+K^-$
from $\pi^+ \pi^-$ and from other intermediate states, in terms of a $K^*$
Regge-trajectory exchange, suggests that $|E/T|\sim\ 0.2$, thus implying
$\b(B^0 \to K^+K^-) \sim 4 \times 10^{-7}$, only an order of magnitude
below the present limit. A future stringent bound on $\b(B^0 \to K^+ K^-)$, at
a level of $10^{-7}$ or lower, would provide a useful limit on rescattering
effects. 

Another way to measure these effects is to compare $B^0 \to K^0 \bar{K}^0$
(given by $P$) with $B^+\to K^+\bar{K}^0$ (given by $P+A$) \cite{GHLR}, both of
which are anticipated to have branching ratios near $10^{-6}$. If rescattering
can be neglected, then $|A/T|\sim f_B/m_B \sim 0.04$, while $|P/T|\sim 0.2$
follows from recent CLEO measurements \cite{etap}. Therefore, the two branching
ratios are expected to be equal within a factor of less than $\simeq 1.5$.
(Electroweak penguins do not affect this relation.) On the other hand,
Regge-model rescattering implies $|A/T|\sim 0.2$, in which case the two
branching ratios may differ substantially, by up to a factor of four or so.
Also, the interference of $P$ and $A$ in $B^+\to K^+\bar{K}^0$ would lead to a
sizable CP asymmetry between the rate of this process and its charge-conjugate.
These measurements could provide useful limits on final state rescattering. 
 
\bigskip

\centerline{\bf V.  GENERALIZATION TO $B \to (K^* \pi,~\rho K)$ DECAYS}
\bigskip

In addition to $B \to K \pi$ decays, one may also obtain information about
$\gamma$ from the analogous decays to a vector meson and a pseudoscalar, $B \to
K \rho$ and $B \to K^* \pi$. Each of these two systems of neutral and charged
$B$ decays involves SU(3) amplitudes of a specific kind \cite{VP}, and can be
studied separately in a way very similar to $B \to K\pi$. The amplitudes of
$B^+ \to K^{*0} \pi^+$ and $B^0 \to K^{*+} \pi^-$ are given in terms of $P'_P$
and $T'_P$, whereas those of $B^+ \to K^0 \rho^+$ and $B^0 \to K^+ \rho^-$ are
described by $P'_V$ and $T'_V$. Here the suffix on each amplitude denotes
whether the spectator quark is included in a pseudoscalar ($P$) or a vector
($V$) meson. 

The values of the hadronic parameters, $|T'/P'|$, $\delta$ and the electroweak
penguin corrections obtain different values in $B \to K \pi,~B \to K^* \pi$ and
$B \to K \rho$. As noted in Sec.~III, the sensitivity of measuring $\gamma$
increases with $|T'/P'|$ since the asymmetry $A_0$, as well as the
deviation of $R$ from 1, are both proportional to this ratio. In Ref.~\cite{VP}
we concluded from recent CLEO data \cite{CLEOVP} on $B^+ \to K^+ \omega$ and
$B^+ \to K^+ \phi$  that $|P'_P| < |P'_V|$. Model-dependent calculations
\cite{Chau,Ciu} predict $|T_P| > |T_V|$. If this turns out to be the case,
namely if $\b(B^+ \to \rho^+ \pi^0) > \b(B^+ \to \rho^0 \pi^+)$ and $\b(B^0 \to
\rho^+\pi^-) > \b(B^0 \to \rho^- \pi^+)$, then one would also conclude that
$|T'_P| > |T'_V|$. Hence, $|T'_P/P'_P| > |T'_V/P'_V|$, implying that in this
respect decays to $K^* \pi$ are more sensitive to a measurement of $\gamma$
than decays to $K \rho$.
\bigskip

\centerline{\bf VI.  CONCLUSIONS}
\bigskip

By measuring the rates for $B^0 \to K^+ \pi^-$, $B^+ \to K^0 \pi^+$, and their
charge-conjugates, it is possible to extract information on the weak phase
$\gamma$.  A useful level of precision ($\Delta \gamma \simeq \pm
10^\circ$) requires about 200 times the present data sample of about a dozen
events in each channel.  This appears to be within the reach of the highest
luminosities attainable at $e^+ e^-$ colliders operating at the $\Upsilon(4S)$,
and may also be feasible in hadronic production of $B$ mesons. 

A key source of uncertainty in the method appears to be the determination of
the magnitude of the tree amplitude ($T'$) interfering with the dominant
penguin amplitude ($P'$).  This requires one to measure the $B^0 \to \pi^+
\pi^-$ or $B^+\to \pi^+\pi^0$ rate to better than 20\% and to make
corresponding improvements in the understanding of how well factorization
applies to the comparison of $B^0 \to\pi^- \ell^+ \nu_\ell$ with the tree
amplitudes of $B^0 \to \pi^+ \pi^-,~ B^+ \to \pi^+\pi^0$ and $B^0 \to K^+
\pi^-$ decays. 

Another theoretical error follows from hadronic uncertainties in calculating
electroweak penguin contributions to the $B \to K \pi$ decay amplitudes. A 5\%
contribution would lead to an uncertainty in $\gamma$ of the order of tens of
degrees. Finally, we argued that an uncertainty at a similar level follows from
possible rescattering effects in $B^+\to K^0\pi^+$.  The importance of such
effects may be found by future measurements of the rates for $B^0 \to K^+
K^-,~B^0 \to K^0 \bar{K}^0$ and $B^+ \to K^+ \bar{K}^0$. 

[Note added:  After the present work was submitted for publication there
appeared a related study \cite{Wurt} involving a slightly broader range of $r =
0.20 \pm 0.07$.]
\bigskip

\centerline{\bf  ACKNOWLEDGEMENTS}
\bigskip

We thank J. Alexander, K. Berkelman, P. Drell, R. Fleischer, L. Gibbons, C.
Roberts, S. Stone and F. W\"urthwein for useful discussions.  M. G. wishes to 
thank the CERN Theory Division for its hospitality. This work was supported in 
part by the United States -- Israel Binational Science Foundation under 
Research Grant Agreement 94-00253/2 and by the United States Department of 
Energy under Contracts No.~DE-FG02-90-ER40560 and DE-FG03-92-ER40701. 
\bigskip

\def \ajp#1#2#3{Am.~J.~Phys.~{\bf#1}, #2 (#3)}
\def \apny#1#2#3{Ann.~Phys.~(N.Y.) {\bf#1}, #2 (#3)}
\def \app#1#2#3{Acta Phys.~Polonica {\bf#1}, #2 (#3)}
\def \arnps#1#2#3{Ann.~Rev.~Nucl.~Part.~Sci.~{\bf#1}, #2 (#3)}
\def \cmp#1#2#3{Commun.~Math.~Phys.~{\bf#1}, #2 (#3)}
\def \cmts#1#2#3{Comments on Nucl.~Part.~Phys.~{\bf#1}, #2 (#3)}
\def \corn93{{\it Lepton and Photon Interactions:  XVI International
Symposium, Ithaca, NY August 1993}, AIP Conference Proceedings No.~302,
ed.~by P. Drell and D. Rubin (AIP, New York, 1994)}
\def \cp89{{\it CP Violation,} edited by C. Jarlskog (World Scientific,
Singapore, 1989)}
\def \dpff{{\it The Fermilab Meeting -- DPF 92} (7th Meeting of the
American Physical Society Division of Particles and Fields), 10--14
November 1992, ed. by C. H. Albright \ite~(World Scientific, Singapore,
1993)}
\def \dpf94{DPF 94 Meeting, Albuquerque, NM, Aug.~2--6, 1994}
\def \efi{Enrico Fermi Institute Report No. EFI}
\def \el#1#2#3{Europhys.~Lett.~{\bf#1}, #2 (#3)}
\def \f79{{\it Proceedings of the 1979 International Symposium on Lepton
and Photon Interactions at High Energies,} Fermilab, August 23-29, 1979,
ed.~by T. B. W. Kirk and H. D. I. Abarbanel (Fermi National Accelerator
Laboratory, Batavia, IL, 1979}
\def \hb87{{\it Proceeding of the 1987 International Symposium on Lepton
and Photon Interactions at High Energies,} Hamburg, 1987, ed.~by W. Bartel
and R. R\"uckl (Nucl. Phys. B, Proc. Suppl., vol. 3) (North-Holland,
Amsterdam, 1988)}
\def \ib{{\it ibid.}~}
\def \ibj#1#2#3{~{\bf#1}, #2 (#3)}
\def \ichep72{{\it Proceedings of the XVI International Conference on High
Energy Physics}, Chicago and Batavia, Illinois, Sept. 6--13, 1972,
edited by J. D. Jackson, A. Roberts, and R. Donaldson (Fermilab, Batavia,
IL, 1972)}
\def \ijmpa#1#2#3{Int.~J.~Mod.~Phys.~A {\bf#1}, #2 (#3)}
\def \ite{{\it et al.}}
\def \jmp#1#2#3{J.~Math.~Phys.~{\bf#1}, #2 (#3)}
\def \jpg#1#2#3{J.~Phys.~G {\bf#1}, #2 (#3)}
\def \lkl87{{\it Selected Topics in Electroweak Interactions} (Proceedings
of the Second Lake Louise Institute on New Frontiers in Particle Physics,
15--21 February, 1987), edited by J. M. Cameron \ite~(World Scientific,
Singapore, 1987)}
\def \ky85{{\it Proceedings of the International Symposium on Lepton and
Photon Interactions at High Energy,} Kyoto, Aug.~19-24, 1985, edited by M.
Konuma and K. Takahashi (Kyoto Univ., Kyoto, 1985)}
\def \mpla#1#2#3{Mod.~Phys.~Lett.~A {\bf#1}, #2 (#3)}
\def \nc#1#2#3{Nuovo Cim.~{\bf#1}, #2 (#3)}
\def \np#1#2#3{Nucl.~Phys.~{\bf#1}, #2 (#3)}
\def \pisma#1#2#3#4{Pis'ma Zh.~Eksp.~Teor.~Fiz.~{\bf#1}, #2 (#3) [JETP
Lett. {\bf#1}, #4 (#3)]}
\def \pl#1#2#3{Phys.~Lett.~{\bf#1}, #2 (#3)}
\def \plb#1#2#3{Phys.~Lett.~B {\bf#1}, #2 (#3)}
\def \pr#1#2#3{Phys.~Rev.~{\bf#1}, #2 (#3)}
\def \pra#1#2#3{Phys.~Rev.~A {\bf#1}, #2 (#3)}
\def \prd#1#2#3{Phys.~Rev.~D {\bf#1}, #2 (#3)}
\def \prl#1#2#3{Phys.~Rev.~Lett.~{\bf#1}, #2 (#3)}
\def \prp#1#2#3{Phys.~Rep.~{\bf#1}, #2 (#3)}
\def \ptp#1#2#3{Prog.~Theor.~Phys.~{\bf#1}, #2 (#3)}
\def \rmp#1#2#3{Rev.~Mod.~Phys.~{\bf#1}, #2 (#3)}
\def \rp#1{~~~~~\ldots\ldots{\rm rp~}{#1}~~~~~}
\def \si90{25th International Conference on High Energy Physics, Singapore,
Aug. 2-8, 1990}
\def \slc87{{\it Proceedings of the Salt Lake City Meeting} (Division of
Particles and Fields, American Physical Society, Salt Lake City, Utah,
1987), ed.~by C. DeTar and J. S. Ball (World Scientific, Singapore, 1987)}
\def \slac89{{\it Proceedings of the XIVth International Symposium on
Lepton and Photon Interactions,} Stanford, California, 1989, edited by M.
Riordan (World Scientific, Singapore, 1990)}
\def \smass82{{\it Proceedings of the 1982 DPF Summer Study on Elementary
Particle Physics and Future Facilities}, Snowmass, Colorado, edited by R.
Donaldson, R. Gustafson, and F. Paige (World Scientific, Singapore, 1982)}
\def \smass90{{\it Research Directions for the Decade} (Proceedings of the
1990 Summer Study on High Energy Physics, June 25 -- July 13, Snowmass,
Colorado), edited by E. L. Berger (World Scientific, Singapore, 1992)}
\def \stone{{\it B Decays}, edited by S. Stone (World Scientific,
Singapore, 1994)}
\def \tasi90{{\it Testing the Standard Model} (Proceedings of the 1990
Theoretical Advanced Study Institute in Elementary Particle Physics,
Boulder, Colorado, 3--27 June, 1990), edited by M. Cveti\v{c} and P.
Langacker (World Scientific, Singapore, 1991)}
\def \yaf#1#2#3#4{Yad.~Fiz.~{\bf#1}, #2 (#3) [Sov.~J.~Nucl.~Phys.~{\bf #1},
#4 (#3)]}
\def \zhetf#1#2#3#4#5#6{Zh.~Eksp.~Teor.~Fiz.~{\bf #1}, #2 (#3) [Sov.~Phys.
- JETP {\bf #4}, #5 (#6)]}
\def \zpc#1#2#3{Zeit.~Phys.~C {\bf#1}, #2 (#3)}

\end{document}